\documentclass[prb,twocolumn,showpacs,preprintnumbers,amsmath,amssymb,floatfix]{revtex4}
\usepackage{graphicx}
\usepackage{dcolumn}
\usepackage{bm}
\usepackage{graphics}
\usepackage{amssymb}
\usepackage{amsmath}
\usepackage{xspace}
\usepackage{epsfig}
\newcommand {\etal}{\begin{itshape}et al\end{itshape}. }

\newcommand {\Sr}{Sr$_2$RuO$_4$ }
\newcommand {\kX}{$\kappa$-(BEDT-TTF)$_2$X }
\newcommand {\kXc}{$\kappa$-(BEDT-TTF)$_2$X, }
\newcommand {\kXs}{$\kappa$-(BEDT-TTF)$_2$X. }
\newcommand {\bX}{$\beta$-(BEDT-TTF)$_2$X }
\newcommand {\bXs}{$\beta$-(BEDT-TTF)$_2$X. }
\newcommand {\bXc}{$\beta$-(BEDT-TTF)$_2$X, }
\newcommand {\bI}{$\beta$-(BEDT-TTF)$_2$I$_3$ }
\newcommand {\kI}{$\kappa$-(BEDT-TTF)$_2$I$_3$ }
\newcommand {\bIBr}{$\beta$-(BEDT-TTF)$_2$IBr$_2$ }
\newcommand {\bIBrc}{$\beta$-(BEDT-TTF)$_2$IBr$_2$, }
\newcommand {\bItBr}{$\beta$-(BEDT-TTF)$_2$I$_2$Br }
\newcommand {\bItBrs}{$\beta$-(BEDT-TTF)$_2$I$_2$Br. }
\newcommand {\bItBrc}{$\beta$-(BEDT-TTF)$_2$I$_2$Br, }

\newcommand {\kCl}{$\kappa$-(BEDT-TTF)$_2$Cu[N(CN)$_2$]Cl }

\newcommand {\Br}{$\kappa$-(BEDT-TTF)$_2$Cu[N(CN)$_2$]Br }
\newcommand {\Brs}{$\kappa$-(BEDT-TTF)$_2$Cu[N(CN)$_2$]Br. }
\newcommand {\Brc}{$\kappa$-(BEDT-TTF)$_2$Cu[N(CN)$_2$]Br, }
\newcommand {\NCS}{$\kappa$-(BEDT-TTF)$_2$Cu(NCS)$_2$ }
\newcommand {\NCSs}{$\kappa$-(BEDT-TTF)$_2$Cu(NCS)$_2$. }
\begin{document}
\title{The Dependence of the Superconducting Transition
Temperature of Organic Molecular Crystals on Intrinsically
Non-Magnetic Disorder: a Signature of either Unconventional
Superconductivity or Novel Local Magnetic Moment Formation}
\author{B. J. Powell\footnote{Electronic address: powell@physics.uq.edu.au} and Ross H. McKenzie}
\affiliation{Department of Physics, University of Queensland,
Brisbane, Queensland 4072, Australia} 

\begin{abstract}
We give a theoretical analysis of published experimental studies
of the effects of impurities and disorder on the superconducting
transition temperature, $T_c$, of the organic molecular crystals
\kX (where X=Cu[N(CN)$_2$]Br and Cu(NCS)$_2$ and BEDT-TTF is
bis(ethylenedithio)tetrathiafulvalene) and $\beta$-(BEDT-TTF)$_2$X
(for X=I$_3$ and IBr$_2$). The Abrikosov--Gorkov (AG) formula
describes the suppression of $T_c$ both by magnetic impurities in
singlet superconductors, including s-wave superconductors and by
non-magnetic impurities in a non-s-wave superconductor. We show
that various sources of disorder (alloying anions,  fast electron
irradiation, disorder accidentally produced during fabrication and
cooling rate induced disorder) lead to the suppression of $T_c$ as
described by the AG formula. This is confirmed by the excellent
fit to the data, the fact that these materials are in the clean
limit and the excellent agreement between the value of the
interlayer hopping integral, $t_\perp$, calculated from this fit
and the value of $t_\perp$ found from angular-dependant
magnetoresistance and quantum oscillation experiments. There are
only two scenarios consistent with the current state of
experimental knowledge. If the disorder induced by all of the four
methods considered in this paper is, as seems most likely,
non-magnetic then the pairing state cannot be s-wave. We show that
published measurements of the cooling rate dependence of the
magnetisation are inconsistent with paramagnetic impurities.
Triplet pairing is ruled out by NMR and upper critical field
experiments. Thus if the disorder is non-magnetic then this
implies that $l\geq2$, in which case Occam's razor suggests that
d-wave pairing is realised in both \bX and \kXs However,
particularly given the proximity of these materials to an
antiferromagnetic Mott transition, it is possible that the
disorder leads to the formation of local magnetic moments via some
novel mechanism. Thus we conclude that either \bX and \kX are
d-wave superconductors or else they display a novel mechanism for
the formation of localised moments, possibly related to the
competition between the antiferromagnetic and superconducting
grounds states. We suggest systematic experiments to differentiate
between these two scenarios.
\end{abstract}

\maketitle

\section{Introduction}\label{sect:intro}

Superconductivity is often found near magnetic ordering. This may
be antiferromagnetic (AFM) order such as in the
cuprates\cite{Alff} and the heavy fermion superconductors
\cite{Mathur} or ferromagnetic order as in the ZrZn$_2$ or UGe$_2$
(see Refs. \onlinecite{Pfleiderer} and \onlinecite{Saxena}
respectively). In each of these cases it is believed that the
superconductivity is unconventional,
\cite{AnnettGoldenfeld&Renn,Agterberg_CeIn3,Kirkpatrick,Ben2} that
is to say that the Cooper pairs have a non-zero angular momentum.
The issue of unconventional superconductivity near magnetic
ordering is of general interest because it may lead to insights
into both non-phononic pairing mechanisms \cite{Bulut} and the
theory of quantum critical points. \cite{Sachdev}

Despite the fact that it is now twenty years since
superconductivity was discovered \cite{Parkin,Yagubskii} in the
layered organic compounds (BEDT-TTF)$_2$X (where BEDT-TTF is
bis(ethylenedithio)tetrathiafulvalene and X is an anion, e.g.,
Cu[N(CN)$_2$]Br or I$_3$) the pairing symmetry remains a matter of
debate. \cite{Lang} BEDT-TTF salts form a number of crystal
structures which are denoted by a Greek letter. All of the crystal
structures consist of alternating layers of BEDT-TTF and an anion.
\cite{Ishiguro} In \bX and \kXc which we consider here, the
BEDT-TTF molecules form a dimerised structure where the anion
removes one electron per dimer. Thus we have alternating
conducting (BEDT-TTF) and insulating (anion) layers. A
particularly interesting feature of these materials is that they
can be driven from an AFM insulating state to a superconducting
state by the application of hydrostatic pressure or by changing
the anion. \cite{Kanoda,Ross_review}

In principle the simplest way to identify the pairing symmetry, or
at least the nodal structure, of a superconductor is to measure
the low temperature behaviour of thermodynamic or transport
properties. For example the specific heat follows an exponentially
activated temperature dependence for a nodeless gap ($C_V \propto
\exp(-|\Delta(0)|/k_BT) $, where $\Delta(0)$ is the
superconducting gap at zero temperature) and a power law
dependence for a gap with nodes ($C_V \propto T^2$ for line nodes
and $C_V \propto T^3$ for point nodes on a three dimensional Fermi
surface). \cite{Mineev&Samokhin} In practice however, there are
difficulties associated with this method of identifying the
pairing symmetry, not the least of which is the need to make
measurements at extremely low temperatures. (Typically a wide
temperature range is required in the region $T/T_c \lesssim 0.2$,
so in the case of $\kappa$-(BEDT-TTF)$_2$Cu[N(CN)$_2$]Br ($T_c\sim
10$~K) one requires measurements taken over a wide range of
temperatures below $\sim 2$~K.) The apparently strong coupling
\cite{Elsinger,Muller} nature of the superconductivity in theses
charge transfer salts means that the behaviour of thermodynamic
and transport functions near $T_c$ is unable to differentiate
between pairing states on symmetry grounds alone and so we must
wait for calculations based on a specific theory of
superconductivity to use this data to examine the pairing
symmetry.

Regardless of the reasons one fact is clear, \cite{Lang} low
temperature behaviours have been, to date, unable to settle the
debate on the pairing symmetry in the layered organic
superconductors. In particular, two pairing symmetries have been
widely discussed: strong coupling s-wave superconductivity and
d-wave pairing.

In \Br the $^{13}$C NMR spin lattice relaxation rate,
\cite{de_Soto,Mayaffre,KanodaNMR} $(T_1)^{-1}$, shows no
Hebel-Slichter peak and a power law cutoff, $(T_1)^{-1} \propto
T^n$, where $n \simeq 3$. A Hebel-Slichter peak is expected for
s-wave pairing while $(T_1)^{-1} \propto T^3$ is expected for line
nodes. \cite{Sigrist&Ueda}

Much controversy has surrounded the London penetration depth with
some groups reporting s-wave pairing
\cite{Lang_penetration2,Lang_penetration3,Dressel,Lang_penetration1,Harshman}
and others finding line nodes consistent with d-wave pairing
\cite{Le1,Le2,Carrington,Pinteric,Achkir,Kanoda_penetration} in
both \Br (Refs.
\onlinecite{Lang_penetration2,Lang_penetration3,Dressel} and
\onlinecite{Le1,Le2,Carrington,Pinteric}) and \NCS (Refs.
\onlinecite{Lang_penetration3,Dressel,Lang_penetration1,Harshman},
\onlinecite{Carrington}, \onlinecite{Le2}, \onlinecite{Achkir} and
\onlinecite{Kanoda_penetration}). However, the most recent
measurements \cite{Carrington} have two advantages over older
experiments. Firstly very low magnetic fields were used. The use
of fields less than the lower critical field is important in
penetration measurements because vortex dynamics are a serious
impediment to accurately measuring the penetration depth. Secondly
Carrington \etal \cite{Carrington} made measurements down to 0.4~K
and therefore made a large range of measurements below $T \sim 0.2
T_c$. This is the lowest temperature range considered in any of
the thermodynamic or transport experiments, making Carrington
$\textit{et al}$.'s conclusions the most reliable drawn from
experiments of this type. Carrington \etal found that the
temperature dependence of the penetration depth of both \Br and
\NCS is inconsistent with a nodeless gap.

Initial measurements of the specific heat of \Br showed a $T^2$
dependence \cite{Nakazawa} but the interpretation of these results
has been questioned. \cite{Lang} More recent measurements of the
specific heat have found an exponentially activated temperature
dependence for both \Br (Ref. \onlinecite{Elsinger}) and \NCS
(Ref. \onlinecite{Muller}).

Several groups have considered probes which do not rely on the low
temperature behaviour of the measurement. Brando \etal
\cite{Bando} and Arai \etal \cite{Aria1,Aria2} attempted to
observe the local density of states (LDOS) of \NCS by measuring
the differential conductance using a scanning tunnelling
microscope (STM). Each of these experiments found a LDOS that is
consistent with d-wave pairing however, none of the experiments
observed the coherence peaks which are a characteristic feature of
the superconducting state and have been observed
\cite{LangSTM,Hudson,Pan} in similar experiments on
Bi$_2$Sr$_2$CaCu$_2$O$_{8+x}$. Also one should note that Bando
\etal \cite{Bando} observed a LDOS in the layered s-wave
superconductor NbN which has the same form as that which is
interpreted as d-wave in experiments on \NCSs

Schrama \etal \cite{Schrama} attempted to determine the anisotropy
in the superconducting order parameter by measuring the
magneto-optical properties of \NCS and found results indicative of
d-wave pairing. However, in light of the debate over the
interpretation of these results
\cite{Hill_comment,Shibauchi,Schrama_reply} one cannot consider
these measurements to have determined the pairing symmetry.

Izawa \etal \cite{Izawa} measured the thermal conductivity tensor
of \NCS in a magnetic field. They observed a four fold anisotropy
at low temperatures which they interpreted as evidence for d-wave
pairing. However, it is possible that the vortices produced in
\NCS are actually Josephson vortices. Therefore it remains to be
shown whether or not the theory
\cite{Kubert&Hirschfeld,Franz,Vekhter&Houghton} on which Izawa
\etal base their analysis is valid for this material.

The $^{13}$C NMR Knight shift has been measured
\cite{de_Soto,Mayaffre} for \Brs With a magnetic field, ${\bf H}$,
parallel to the conducting planes, as $T \rightarrow 0$ so does
the Knight shift.\footnote{Mayaffre {\textit{et al}}.
\cite{Mayaffre} also reported the Knight shift for ${\bf H}$
perpendicular to the conducting planes, but for ${\bf
H}>H_{c2}^a$, where $H_{c2}^a$ is the upper critical field with
the field perpendicular to the conducting planes.} This does not
actually rule out triplet pairing, although it does make triplet
pairing extremely unlikely. This experiment is compatible with a
triplet state in which ${\bf d}({\bf k})\times{\bf H} = 0$ where
${\bf d}({\bf k})$ is the usual Balian--Werthamer order parameter
for triplet superconductivity. \cite{Balian&Werthamer,Vollhardt}
An example of a triplet phase compatible \cite{Ben1,Ben3} with
this experiment is an A-phase with ${\bf d}({\bf k})$ pinned to
the c-axis, \footnote{For \kX the $b$-$c$ plane is the conducting
plane and the $a$-axis is perpendicular to the conducting planes.}
which is not an impossibility given the highly anisotropic nature
of \Brs However, Zuo \etal \cite{Zuo} measured the critical field
as a function of temperature with ${\bf H}$ parallel to the
conducting planes. In this configuration no orbital currents flow
so the critical field is due to Clogston--Chandrasekhar (or Pauli)
limit. \cite{Clogston,Chandrasekhar,Ben3} There is no
Clogston--Chandrasekhar limit for ${\bf H} \perp c$ for triplet
states compatible with measured Knight shift. Thus for such states
there would be no critical field with ${\bf H} \| b$ (in fact for
such states one would increase $T_c$ by applying a field parallel
to the b-axis \cite{Ben3}). Experimentally \cite{Murata} it is
found that superconductivity is destroyed by a magnetic field
parallel to the b-axis. Therefore only when considered together do
the three experiments discussed above \cite{de_Soto,Zuo,Murata}
strictly rule out triplet pairing. \footnote{While these
experiments rule out triplet pairing, they do not strictly rule
out pairing with an odd angular momentum. The total wavefunction
of a pair of fermions must be antisymmetric under the exchange of
all labels. Usually one only considers the spin and momentum
labels. However, it is also possible to construct order parameters
that are antisymmetric under the operation $\omega_n \rightarrow
-\omega_n$. The possibility of \lq odd frequency' s-wave triplet
pairing has been consider in the context of both superfluid $^3$He
(Ref. \onlinecite{Berezinskii}) and the cuprates.
\cite{Balatsky&Abrahams} Although it is not thought that odd
frequency pairing is realised in either of these systems there is
no reason on symmetry grounds to exclude odd frequency pairing in
the layered organics. For example, singlet, p-wave, odd frequency
pairing is compatible with all of the experiments discussed in
this paper. However, as there are no known examples of odd
frequency pairing and no evidence of odd frequency pairing in
either \bX or \kX we will not consider this possibility further
here. The possibility of s-wave, triplet, odd frequency pairing
has been discussed in the context of \kX by Vojta and Dagotto,
\cite{Vojta} who considered triplet, s-wave pairing. Odd
frequency, triplet, s-wave pairing is insensitive to non-magnetic
disorder for the same reasons as even frequency, singlet, s-wave
pairing is. \cite{Abrahams}} Further evidence for
Clogston--Chandrasekhar limiting comes from the observation that
the in plane upper critical field is independent of the field
direction. \cite{Nam} Given the anisotropic nature of the Fermi
surface of \Br it is extremely unlikely that orbital mechanisms
for the destruction of superconductivity would be so isotropic.

The results of quantum chemistry calculations suggest that the
simplest theoretical model which can describe these materials is a
half-filled Hubbard model on an anisotropic triangular lattice.
\cite{Kino&Fukuyama,Ross_review} Because of the proximity of the
antiferromagnetic insulating phase and the superconducting phase
several groups have examined the possibility of spin fluctuation
induced superconductivity within the confines of this model using
a variety of techniques, including mean-field theory,
\cite{Kino&Fukuyama} the fluctuation-exchange approximation,
\cite{Schmalian,Kino&Kontani,Kondo&Moriya} third order
perturbation theory, \cite{Jujo} weak coupling renormalisation
group analysis, \cite{Tsai&Marston} the random phase approximation
\cite{Louati,Vojta} and quantum Monte Carlo methods.
\cite{Kuroki&Aoki} All of these groups concluded that spin
fluctuations lead to d-wave pairing. These authors found an
enhanced dynamical susceptibility at $(\pi,\pm\pi)$ which leads to
d$_{x^2-y^2}$ pairing. Alternatively, both d-wave
\cite{Varelogiannis} and s-wave \cite{Girlando1,Girlando2,Pedron}
pairing symmetries have been considered in the context of phononic
pairing mechanisms.

So, perhaps the only emerging consensus is that the low
temperature behaviours have not been able to conclusively settle
the debate between s-wave and d-wave pairing symmetries. In the
remainder of this paper we will investigate how the effects of
disorder can be used to distinguish between these two symmetries.

\section{The Abrikosov--Gorkov formula}\label{sect:AG}

Anderson's theorem \cite{Andersons_theorem} states that for s-wave
pairing non-magnetic impurities do not change $T_c$. This is
because Cooper pairs are formed from time reversed states and
although non-magnetic impurities may change, for example, the
phonon spectrum, they do not break time reversal symmetry (TRS).
However, magnetic impurities strongly reduce $T_c$ for all singlet
states because they do break TRS. \cite{Maple} This behaviour is
described by the Abrikosov--Gorkov (hereafter AG) formula:
\cite{AG}
\begin{eqnarray}
\ln \left(\frac{T_{c0}}{T_{c}} \right) = \psi\left( \frac{1}{2} +
\frac{\hbar}{4\pi k_BT_{c}}\frac{1}{\tau_{M}} \right) - \psi\left(
\frac{1}{2} \right) \label{eqn:AGM}
\end{eqnarray}
where $T_{c0}$ is the superconducting critical temperature in the
pure system and $\psi(x)$ is the digamma function. $\tau_{M}$ is
the quasiparticle lifetime due to scattering from magnetic
impurities. Assuming \emph{isotropic} scattering $\tau_M$ is given
by \cite{Rickayzen}
\begin{eqnarray}
\frac{\hbar}{\tau_M} = N_M\pi J_i(J_i+1)N(0)|u_M|^2
\label{eqn:mag_lifetime}
\end{eqnarray}
where $N_M$ is the number density of magnetic impurities, $N(0)$
is the density of states per spin at the Fermi level, $J_i$ is the
total angular momentum of the paramagnetic atoms and $u_M$ is the
amplitude for scattering from a magnetic impurity.

In the superconducting state the anomalous Green's function,
$F_{\alpha\beta}({\bf k}, \omega_n)$, is finite and therefore
there is, in the presence of non-magnetic impurities, an anomalous
self energy, $\Sigma_{2,\alpha\beta}(\omega_n)$, which, in $n$
dimensions, is given by \cite{Mineev&Samokhin}
\begin{eqnarray}
\Sigma_{2,\alpha\beta}(\omega_n) = \frac{1}{2\pi N(0)\tau_N}
\int{\frac{d^nk}{(2\pi)^n}F_{\alpha\beta}({\bf k}, \omega_n)}
\label{eqn:anom_self_energy}
\end{eqnarray}
where $\tau_N$, the lifetime for scattering from non-magnetic
impurities is given by  \cite{Rickayzen}
\begin{eqnarray}
\frac{\hbar}{\tau_N} = N_N\pi N(0)|u_N|^2
\label{eqn:non_mag_lifetime}
\end{eqnarray}
where $N_N$ is the number density of non-magnetic impurities and
$u_N$ is the amplitude for scattering from a non-magnetic
impurity.

For s-wave pairing $\Sigma_{2,\alpha\beta}(\omega_n)$ is clearly
finite, and it can be shown that the anomalous self energy cancels
exactly with the normal self-energy,
$\Sigma_{1,\alpha\beta}(\omega_n)$, when the critical temperature
is evaluated. Therefore $T_c$ is unchanged by non-magnetic
impurities for an s-wave superconductor, as expected from
Anderson's theorem. \cite{Andersons_theorem} However, for
non-s-wave pairing \footnote{For certain of Fermi surfaces
non-s-wave pairing can include extended s-wave pairing. The
important feature of what we term a non-s-wave is that the phase
of the order parameter changes sign around the Fermi surface.
Clearly this can be the case for extended s-wave states on an
anisotropic Fermi surface.} it can be seen, from symmetry grounds
alone, that the integral in equation (\ref{eqn:anom_self_energy})
vanishes. Thus the anomalous self-energy does not cancel the
normal self energy and $T_c$ is lowered by non-magnetic impurities
in a non-s-wave superconductor. Further, it can be shown that for
pairing states with non-s-wave symmetry non-magnetic impurities
reduce $T_c$ via the Abrikosov--Gorkov formula.
\cite{Larkin,Mineev&Samokhin} However, in this case
\begin{eqnarray}
\ln \left(\frac{T_{c0}}{T_{c}} \right) = \psi\left( \frac{1}{2} +
\frac{\hbar}{4\pi k_BT_{c}}\frac{1}{\tau_{N}} \right) - \psi\left(
\frac{1}{2} \right) \label{eqn:AGNM}
\end{eqnarray}
where again we have assumed isotropic scattering. The
predictions\footnote{For very large levels of non-magnetic
impurities a change in $T_c$ is predicted, and indeed observed,
for s-wave superconductors. The two main points to note about this
change in $T_c$ are: (i) the number of impurities required to
produce a small change in $T_c$ in an s-wave superconductor would
be enough to completely suppress superconductivity in a non-s-wave
superconductor and (ii) the change in $T_c$ due to non-magnetic
impurities in a s-wave superconductor is not described by the AG
formula (see section \ref{sect:AG}). Indeed non-magnetic
impurities can both decrease and increase $T_c$ in an s-wave
superconductor. In this paper however we will only consider low
concentrations of impurities. At these levels of impurities $T_c$
of an s-wave superconductor is unchanged by non-magnetic
impurities to an extremely good approximation.} of Anderson's
theorem have been confirmed for the alloys of many s-wave
superconductors. \cite{Caroli,Markowitz,Tsuneto,Lynton}

Hasegawa and Fukuyama \cite{Hasegawa&Fukuyama} suggested that weak
localisation could lead to an alternative mechanism for the
suppression of $T_c$ in organic superconductors. Notably this
mechanism allows for the suppression of $T_c$ by non-magnetic
disorder in s-wave superconductors, in violation of Anderson's
theorem. However the Hasegawa--Fukuyama mechanism has a
dramatically different $\tau$ dependence to the AG formula. We
will show in this paper that the observed suppression of $T_c$ in
\bX and \kX is described by the AG formula and therefore the
predictions of Hasegawa and Fukuyama are not in agreement with
experiment. For a multiband superconductor interband scattering
processes can also lead to a suppression in $T_c$ (see, for
example Ref. \onlinecite{Golubov}). However, of the two polymorphs
discussed in this paper only one ($\kappa$-(BEDT-TTF)$_2$X) has
multiple sheets to its Fermi surface. As it seems reasonable to
assume (unless evidence is found to the contrary) that the
suppression of $T_c$ in both materials is due to the same
mechanism we will not discuss interband scattering effects
further. Also not that for moderate amounts of disorder interband
scattering effects and the AG formula give very different
predictions for the suppression of $T_c$.

It can be shown that the digamma function has the property
\begin{eqnarray}
\psi\left(\frac{1}{2}+x\right) = \psi\left(\frac{1}{2}\right) +
\frac{\pi^2x}{2} + {\cal{O}}(x^2).
\end{eqnarray}
Hence for $\hbar/\tau \ll k_BT_c$ (i.e. as the number of
impurities tends to zero) the AG equation becomes
\begin{eqnarray}
T_{c0} - T_c \simeq \frac{\pi\hbar}{8k_B}\frac{1}{\tau}.
\label{eqn:linearisedAG}
\end{eqnarray}
Clearly the above is valid for both magnetic impurities in singlet
states ($\tau = \tau_M$) and non-magnetic impurities in non-s-wave
pairing states (in which case $\tau = \tau_N$).

\subsection{Mixed order parameters}

As well as s-wave pairing and non-s-wave pairing, a third logical
possibility exists: a state which contains a superposition of both
s- and non-s-wave pairing. For example the
$\textrm{s}+i\textrm{d}$ and $\textrm{s}+\textrm{d}$ states. In
general such a state can be written as
\begin{eqnarray}
\Delta(\tau)=\Delta_0(\tau)\left(\cos\left(\varphi(\tau)\right)\hat\Delta_s
+ e^{i\theta}\sin\left(\varphi(\tau)\right)\hat\Delta_n\right)
\end{eqnarray}
where $\tau$ is the quasiparticle lifetime, $\Delta(\tau)$ is the
order parameter of the superconductor, $\Delta_0(\tau)$ gives the
magnitude of the order parameter, $\hat\Delta_s$ is a function
with a magnitude of unity and s-wave symmetry, $\hat\Delta_n$ is a
function with a magnitude of unity and the appropriate non-s-wave
symmetry and $\theta$ and $\varphi(\tau)$ parameterise the
superposition. For clarity we have suppressed all spin and
momentum labels. We will describe this state as the
$\textrm{s}+\textrm{n}$ state.

Na\"ively, it might appear that the $\textrm{s}+\textrm{n}$ state
might explain the low temperature behaviour of the thermodynamic
and transport properties. If the states had a large d-wave
component it would appear to have nodes at high temperatures, but
at low temperatures the small fully gapped s-wave part of the
order parameter would cause an exponential cutoff. However, a more
careful analysis of the data shows that this scenario is not what
has been observed, indeed the results of the experiments performed
to the lowest temperatures suggested nodes in the gap.
\cite{Carrington}

To describe the effect of disorder on the $\textrm{s}+\textrm{n}$
state we will begin my studying the two extreme cases of total
coherence between the states and zero coherence between the
states. It will then be seen that all other possibilities are
intermediates of these two extremes.

If there is total coherence between the states, then adding
disorder does not change the ratio between the s-wave and
non-s-wave parts of the order parameter, i.e. $\varphi$ is
independent of $\tau$. It is straightforward to show that, subject
to this constraint,
\begin{eqnarray}
\ln\left(\frac{T_{c0}}{T_c}\right) = 2\pi N(0)VT \sum_{n\geq0}
\left(\frac{R}{\omega_n+1/2\tau} - \frac{1}{\omega_n}\right).
\end{eqnarray}
where $V$ is the effective pairwise interaction between the
electrons. For s-wave pairing in the presence of non-magnetic
impurities \cite{AGD}
\begin{eqnarray}
R=1+\frac{1}{2\tau|\omega_n|}
\end{eqnarray}
and one finds that $T_c=T_{c0}$ independent of $\tau$, in
confirmation of Anderson's theorem. But, for non-s-wave pairing
$R=1$ and we arrive at the AG equation (\ref{eqn:AGNM}).

For an $\textrm{s}+\textrm{n}$ superconductor
\begin{eqnarray}
R=1+\frac{\alpha(\varphi)}{2\tau|\omega_n|}.
\label{eqn:R_alpha_phi}
\end{eqnarray}
$\alpha(\varphi)$ is an unknown function, however it is clear that
$\alpha(0) = 1$ and $\alpha(\pi) = 0$. Thus one finds that
\begin{eqnarray}
\ln \left(\frac{T_{c0}}{T_{c}} \right) = \psi\left( \frac{1}{2} +
\frac{\hbar}{4\pi k_BT_{c}}\frac{1}{(1-\alpha)\tau} \right) -
\psi\left( \frac{1}{2} \right) \label{eqn:AGsin}
\end{eqnarray}
Thus we find that rigid coherence in an $\textrm{s}+\textrm{n}$
superconductor simply \lq renormalises' the quasiparticle lifetime
in the AG equation.

For a superconductor without coherence between the two parts of
the order parameter $\varphi$ varies strongly with $\tau$ and the
two parts of the order parameter are independent of one another.
Thus non-magnetic disorder does not change the bulk critical
temperature because of the s-wave part of the wavefunction. But
non-magnetic disorder would reduce the critical temperature for
the non-s-wave part of the wavefunction. This would lead to there
being two phase transitions in the presence of non-magnetic
disorder, the first from the non-superconducting state to an
s-wave superconductor and the second from an s-wave superconductor
to an $\textrm{s}+\textrm{n}$ superconductor. Two such phase
transitions would have a clear experimental signature. For example
there would be two anomalies is the specific heat. This has, to
the best of our knowledge, never been observed in the layered
organic superconductors. Therefore we can rule out the possibility
of $\textrm{s}+\textrm{n}$ superconductivity with zero or, indeed,
weak coherence between the states on phenomenological grounds.

\subsection{Non-magnetic disorder in other superconductors}\label{sect:other_SCs}

The effects of non-magnetic disorder have been carefully observed
in several other superconductors. The best known case is
Sr$_2$RuO$_4$. Mackenzie \etal \cite{Mackenzie_Tc} measured $T_c$
for several samples with varying residual resistivities. Assuming
the Drude model of conductivity they found the variation of $T_c$
with $\rho_0$ to be in excellent agreement with the AG formula.

Both magnetic (Ni) and intrinsically non-magnetic (Zn, Pr, fast
electron irradiation) defects lead to the suppression of $T_c$ of
YBaCu$_3$O$_{6+x}$ (YBCO) in line with the AG formula.
\cite{Tolpygo1,Tolpygo2} However, it is known \cite{Mahajan} that
the substitution of Zn atoms for Cu atoms in the CuO$_2$ planes of
YBCO can lead to the formation of localised magnetic moments. It
is thought that these local moments form on the nearest neighbour
Cu atoms rather than on the Zn site itself. \cite{Mahajan} There
has been much debate \cite{RAAT01,RAAT03} as to whether the
mechanism for pair breaking in YBCO crystals with Zn impurities is
local moment scattering or potential scattering due to the Zn
impurity (of course the two mechanisms are not mutually exclusive
\cite{Pint&Schachinger}). Recent work by Davis \etal
\cite{Pan,Hudson} indicates that non-magnetic scattering is the
dominant mechanism by which Zn impurities \cite{Pan} lower $T_c$
and further that even the magnetic impurities (Ni) act primarily
as potential scatterers. \cite{Hudson}

In the heavy fermion superconductor UPt$_3$ a suppression of $T_c$
has been observed that is consistent with the AG theory.
\cite{Dalichaouch_AG_UPt3,Duijn} Surprisingly both magnetic (Ni)
impurities and non-magnetic (Gd) impurities suppress $T_c$ in the
same way. \cite{Duijn} In light of the discovery that Ni
impurities act primarily as potential scatterers in YBCO it seems
plausible that the same thing may happen in UPt$_3$. Alternatively
some unknown mechanism may be inducing local moments around the Gd
atoms. This seems unlikely as for this to be consistent with the
observation that Gd and Ni impurities suppress $T_c$ in the same
way this scenario would require the moment induced around Gd atoms
to be the same as the moment due to Ni atoms.

The Bechgaard salts, (TMTSF)$_2$X where (TMTSF is
tetramethylteselanafulvalene and X is an anion, for example
ClO$_4$ or ReO$_4$) are also very sensitive to non-magnetic
disorder. It has been suggested that this is because they are
quasi-one-dimensional systems.
\cite{Tsobnang,Hasegawa&Fukuyama,Ishiguro,MYChoi} Disorder can be
induced by x-ray irradiation, alloying or by a cooling rate
controlled anion disorder transition (which we will discuss
further below). All of these sources of disorder can reduce $T_c$
and can even suppress superconductivity altogether and lead to the
formation of a spin density wave. \cite{Tsobnang,Zuppiroli}

\section{\bX}

There are a series of competing ground states in both \bX and \kX
including antiferromagnetism and superconductivity. By applying
pressure or changing the anion the ground state of these layered
organic crystals can be changed, thus it is thought that different
anions apply different \lq chemical pressures'.
\cite{Kanoda,Ross_review} For superconducting crystals pressure
lowers $T_c$. Thus one might expect that by alloying anions one
could observe the same change in $T_c$ due to the change in \lq
chemical pressure'. However, if one adds small amounts of a second
anion the second anion sites will act as non-magnetic impurities.
Thus, unless the pairing state is s-wave, alloying anions will
suppress $T_c$. The suppression of $T_c$ should be governed by the
AG formula.

Tokumoto \etal \cite{Tokumoto} have produced alloys in the series
$\beta$-(BEDT-TTF)$_2$(I$_3$)$_{1-x}$(IBr$_2$)$_x$. For $x=0$ they
found that $T_c=7.4$~K and for $x=1$ they found $T_c=2.4$~K. Based
on Anderson's theorem one expects that for s-wave pairing $T_c$
will vary monotonically with $x$. However, Tokumoto \etal found no
indications of superconductivity for $0.2 \lesssim x \lesssim
0.7$. A natural explanation of this experiment is that for small,
non-zero values of $x$ the IBr$_2$ anions act as (intrinsically)
non-magnetic impurities in \bI and thus quickly reduce $T_c$ to
zero. Similarly for $x\lesssim 1$ the I$_3$ anions act as
impurities in \bIBr and reduce $T_c$ to zero for quite small
concentrations. This explanation of course requires non-s-wave
pairing.

\begin{figure}
    \centering
    \epsfig{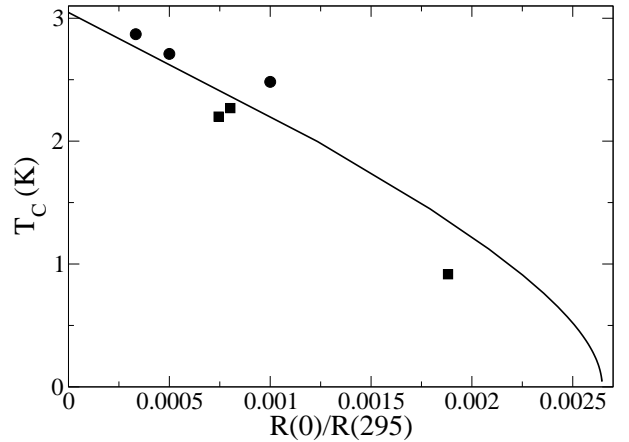}
    \caption{The variation of the superconducting transition
    temperature of \bIBr with the residual resistance ratio, $R(0)/R(295)$.
    The curve is
    a fit, using the AG formula assuming the residual resistivity,
    $\rho_0 \propto 1/\tau_t$, where $\tau_t$ is the quasiparticle lifetime, to the data of Tokumoto $\textit{et al}$.
    \cite{Tokumoto} (squares)
    who induced disorder by substituting I$_3$ anions for IBr$_2$
    and Shegolev and
    Yagubskii \cite{Shegolev} (circles) who reported resistivity
    measurements for several samples. This indicates that either both
    types of impurity induce magnetic moments or else the pairing symmetry is
    non-s-wave. Note that although we have written $R(0)$ Tokumoto $\textit{et al}$.
    did not actually report $R(0)/R(295)$, but $R(T_c)/R(295)$
    thus their data (squares) should be shifted slightly to the left.
    As Shegolev and Yagubskii reported $R(T)/R(295)$ for a range of
    temperatures near $T_c$ we were able to fit to their data to the form
    $R(T)/R(295)=R(0)/R(295)+AT^2$ and thus
    determine both $R(0)/R(295)$ and $T_c$ accurately.} \label{figure:IBr2}
\end{figure}

In figure \ref{figure:IBr2} we plot the data for $T_c$ against
$\rho_0$ for $\beta$-(BEDT-TTF)$_2$(I$_3$)$_{1-x}$(IBr$_2$)$_x$
with $x \lesssim 1$ from Tokumoto \etal \cite{Tokumoto} on the
same graph as data for $\beta$-(BEDT-TTF)IBr$_2$ samples
\cite{Shegolev} which have differing residual resistivities
because of impurities accidently induced in the fabrication
process. The excellent agreement with the AG formula is strong
evidence against the weak localisation scenario. In this fit we
assume only that $\rho_0 \propto 1/\tau_N$. There were not enough
data points reported for $x \lesssim 0$ to make a similar
comparison for $\beta$-(BEDT-TTF)I$_3$. For a more detailed
discussion of the role of disorder in $\beta$-(BEDT-TTF)I$_3$ see
Ref. \onlinecite{betaHbetaL}.

It is also interesting to note that the compound \bItBr is not
superconducting. For \bXc when X is a trihalide, the three
positions of the halide atoms are crystallographically distinct.
In \bI the three iodine atoms are arranged approximately linearly
(which we represent by I-I-I) and are clearly indistinguishable
particles. In \bIBr the atoms are arranged Br-I-Br, that is to say
that the iodine atom is always in one particular location. But, in
\bItBrc the atoms can either be arranged I-I-Br or Br-I-I. This
means that the crystal is intrinsically disordered. \bItBr is
found to have a high residual resistivity. \cite{Tokumoto} Thus we
propose that it is the intrinsically non-magnetic disorder, caused
by the two possible arrangements of the anion, that suppresses
superconductivity in \bItBrs Further Tokumoto \etal observed that
no samples with $R(0)/R(295)\gtrsim0.3$ from any of the alloys
$\beta$-(BEDT-TTF)$_2$(I$_3$)$_{1-x}$(IBr$_2$)$_x$,
$\beta$-(BEDT-TTF)$_2$(IBr$_2$)$_{1-x}$(I$_2$Br)$_x$ or
$\beta$-(BEDT-TTF)$_2$(I$_2$Br)$_{1-x}$(I$_3$)$_x$ superconducted.
This is exactly what one would expect from the AG formalism (c.f.
figure \ref{figure:IBr2}).

At this stage it may appear that the arguments presented above are
in contradiction to what is known about the cuprate
superconductors. These materials have d-wave order parameters and
yet non-stoichiometric compounds often have far higher transition
temperatures than the (stoichiometric) parent compounds (indeed in
many cases the parent compound is non-superconducting). An
excellent example of this is La$_{2-x}$Sr$_x$CuO$_4$ for which
optimal doping is $x\sim0.15$. It was suggested
\cite{Hirschfeld&Goldenfeld} that d-wave superconductivity is
observed in non-stoichiometric compounds because the Born
approximation is not valid for the cuprates. However, it has been
shown \cite{Radtke} that even in the unitary (or resonant)
scattering limit which is appropriate for the cuprates
non-magnetic disorder still destroys d-wave pairing in line with
the predictions of the AG formula and leaves s-wave pairing
unaffected. Further unitary scattering is the appropriate limit
\cite{Pethick} for the unconventional superconductor \cite{Sauls}
UPt$_3$ and in this material $T_c$ is suppressed by non-magnetic
impurities in a manner consistent with the AG formula \cite{Duijn}
as discussed in section \ref{sect:other_SCs}.

However, sofar we have neglected the major difference between
non-stoichiometric compounds in the organics and the cuprates. In
the cuprates the change in stoichiometry introduces a change in
the current carrier concentration. This dramatically alters the
ground state of the cuprates, this effect is absent in the
organics because all of the anions have the same
electronegativity. It should be noted however that, both the
cuprates and the organics are similarly two dimensional as is
attested by the ratio of the zero temperature interlayer coherence
length, $\xi_\perp(0)$, to the interlayer spacing, $a$. For
example in \Br (Ref. \onlinecite{Ishiguro}) $\xi_\perp(0)/a =
5.8/30.016 = 0.19$ and in the cuprates
\cite{Hussey_cuprates_coherence_length} $\xi_\perp(0)/a \sim
0.06-0.45$. Therefore, as both compounds are quasi-two-dimensional
and alloying anions suppresses $T_c$ in \bXc it cannot be merely
the two dimensional nature of the cuprates which is responsible
for observation of superconductivity in non-stoichiometric
compounds.

It has been shown \cite{Tokumoto} that by alloying anions one can
introduce enough disorder into the system to suppress
superconductivity. Assuming that this disorder is non-magnetic
this rules out $\textrm{s}+\textrm{n}$ superconductivity with
anything other than completely rigid coherence between the two
states (that is to say that $\alpha$ is independent of $\varphi$
in the language of equation (\ref{eqn:R_alpha_phi})). Any other
type of coherence would leave a small residual s-wave component
even in the presence of very large amounts of disorder.

\begin{figure}
    \centering
    \epsfig{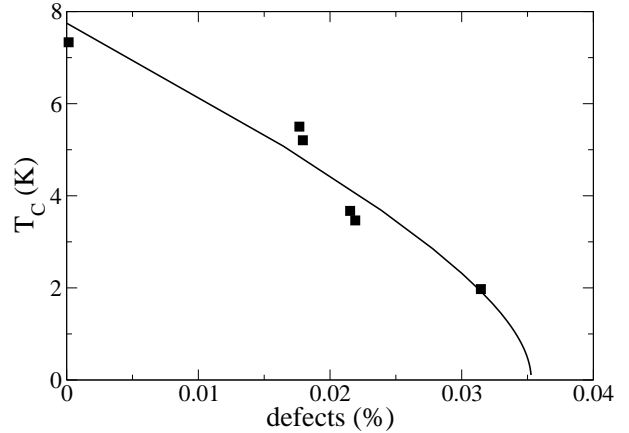}
    \caption{The variation of the superconducting transition
    temperature of \bI with the number of impurities. The data is taken from
    Forro $\textit{et al}$. \cite{Forro}
    who induced defects by irradiating samples with fast
    electrons. The curve is a fit to the AG formula and equation (\ref{eqn:non_mag_lifetime}).
    This indicates that either
    the radiation induces magnetic moments or else the pairing symmetry is
    non-s-wave.} \label{figure:I3}
\end{figure}

Defects can also be induced in materials by irradiating them with
fast electrons. \cite{Tolpygo2,Kirk} Such experiments were
performed on \bI by Forro \etal \cite{Forro} who noted a marked
drop in $T_c$ as the number of defects increased. From figure
\ref{figure:I3} it can be seen that the fit to the AG formula and
equation (\ref{eqn:non_mag_lifetime}) is excellent. Unfortunately
Forro \etal did not report the residual resistivity of their
irradiated samples so a comparison with transport theory cannot be
made. Again the excellent fit of the data to the AG theory is
strong evidence against the weak localisation theory.

We have therefore shown that impurities in \bX suppress $T_c$ via
the AG mechanism for three sources of impurities: alloying anions,
fast electron irradiation and accidentally created defects from
the fabrication process. There is no obvious mechanism for any of
these methods to form magnetic scattering centres. Thus the most
natural interpretation is that there is non-s-wave pairing in \bX
and the reduction in $T_c$ is due to potential scattering.
However, there is a strong similarity between the layered organic
superconductors and the cuprates
\cite{Kino&Fukuyama,Ross_review,Ross_science,Kanoda} in particular
both are close to an antiferromagnetic phase. As we have already
noted, the substitution of Zn for Cu in the CuO$_2$ planes of YBCO
leads to the unexplained formation of local moments on the Cu
atoms neighbouring the Zn impurity. Therefore one must consider
the possibility that a novel mechanism is creating local moments
in all three of experiments discussed above. This may seem
unlikely, but until further experimental evidence on the nature of
the impurities formed in these experiments becomes available we
cannot use disorder to unambiguously determine whether or not
there is s-wave pairing in \bXs

\section{\kX}

One of the most unusual features of \Br is that $T_c$ is dependent
on the rate at which the sample is cooled from $T \gtrsim 80$~K
(Ref. \onlinecite{Su,Stalcup}). The residual resistivity along the
c-axis, $\rho_0$, is also dependent on the cooling rate. It would
appear then, that if one cools \Br quickly one can \lq freeze in'
disorder, whereas if the cooling is slower then the disorder can
relax out. The observation that this disorder suppresses $T_c$
implies that if the pairing state has s-wave symmetry then the
disorder must arise from magnetic impurities, but if another
pairing symmetry is realised then this disorder may arise from
non-magnetic impurities.

There is always a certain amount of intrinsically non-magnetic
impurities in any given crystal. These \lq structural' impurities
will also contribute to the residual resistivity, but they only
effect $T_c$ in the non-s-wave case. We denote the quasiparticle
lifetime caused by this structural disorder by $\tau_s$. Similarly
we will denote the quasiparticle lifetime caused by the cooling
rate induced disorder by $\tau_c$.

\begin{figure}
    \centering
    \epsfig{figure=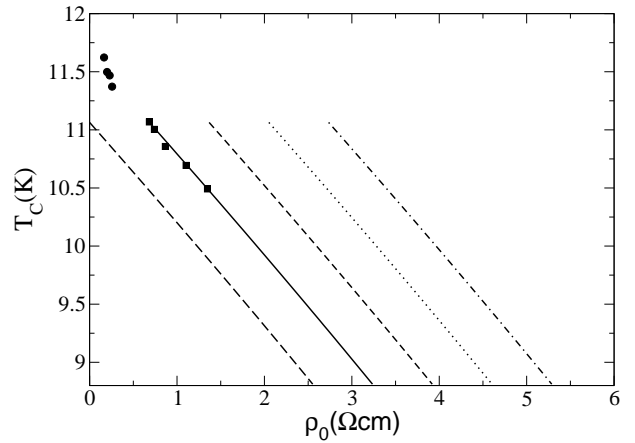, width=6.5cm, angle=270}
    \caption{Variation of the superconducting transition temperature,
    $T_c$, of \Br with the interlayer residual resistivity, $\rho_0$.
    The solid line is a fit to the
    data of Su \emph{et al.} \cite{Su}
    (squares). The other lines are predictions of the s-wave theory for other
    samples with different amounts of structural disorder and
    thus a different $\tau_s$. This structural disorder is assumed
    to be non-magnetic. Thus for s-wave pairing the structural
    disorder changes $\rho_0$ but does not affect $T_c$.
    The data of Stalcup \emph{et al.} \cite{Stalcup} then represents a
    test of the theory. It can clearly be seen that the
    theory does not describe the data as both $T_c$ and $\rho_0$
    are changed for all cooling rates. This indicates the $\tau_s$
    is different for both samples and therefore that either the assumption
    of non-magnetic structural disorder is incorrect or the
    assumption of s-wave pairing is incorrect.
    Note that we have reanalysed the experimental data and used a consistent definition
    of both $T_c$ (based on when the resistivity falls to half of
    its normal state value) and $\rho_0$ (based on a fit to the
    form $\rho(T) = \rho_0 + AT^2$; Matthiessen\rq s rule\cite{Ashcroft&Mermin} was found to be
    obeyed).} \label{fig:s-wave_T_c_expt}
\end{figure}

\begin{figure}
    \centering
    \epsfig{figure=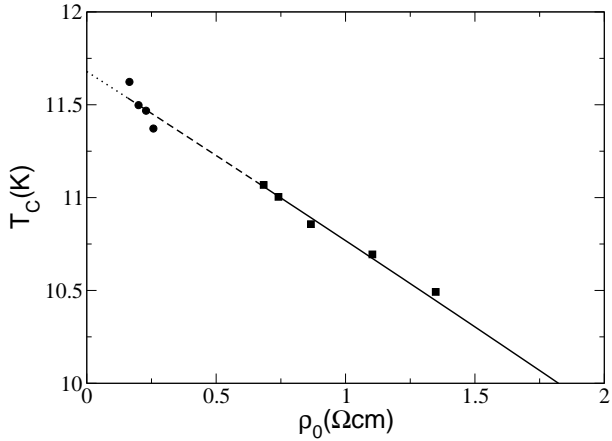, width=6.5cm, angle=270}
    \caption{Variation of the superconducting transition temperature,
    $T_c$, of \Br with the interlayer residual resistivity, $\rho_0$.
    The solid line is a fit to the
    data of Su \emph{et al.} \cite{Su}
    (squares). For non-s-wave pairing the structural
    disorder both changes $\rho_0$ and $T_c$.
    The data of Stalcup \emph{et al.} \cite{Stalcup} then represents a
    test of the theory. The broken lines are a prediction of the non-s-wave theory for other
    samples with different levels of structural disorder and
    thus a different $\tau_s$. It can clearly be seen that the
    theory describes the data as both $T_c$ and $\rho_0$
    are changed for all cooling rates in line with the predictions of
    the AG formula and equation (\ref{eqn:interlayer_resistivity}). The
    dashed portion of the line describes the data of Stalcup \emph{et al.},
    the dotted line is the prediction for a crystal with even less structural disorder.
    The experimental data and the solid line are identical to those
    shown in figure \ref{fig:s-wave_T_c_expt}. Note however that
    this figure also represents the prediction for s-wave pairing
    assuming that the structural impurities are solely magnetic scatterers.} \label{fig:d-wave_T_c_expt}
\end{figure}

As non-magnetic impurities do not affect $T_c$ for s-wave pairing
$T_c$ is given by (\ref{eqn:AGM}) with $\tau_M=\tau_c$. On the
other hand both scattering from magnetic and non-magnetic
impurities contribute to  the residual resistivity so we might
expect
\begin{eqnarray}
\rho_0 \propto \frac{1}{\tau_t} \equiv \frac{1}{\tau_s} +
\frac{1}{\tau_c} \label{eqn:resistivity_propto}
\end{eqnarray}
where $\tau_t$ is the appropriate quasiparticle lifetime for
transport experiments.

The fabrication of different samples will lead to different values
of $\tau_s$. For s-wave pairing this will cause a variation in
$\rho_0$ but not $T_c$ thus one reaches the conclusion that
different samples cooled at the same rate will have different
residual resistivities, but the same maximum critical temperature.
In figure \ref{fig:s-wave_T_c_expt} we fit the linearised AG
equation (\ref{eqn:linearisedAG}) to the data of Su \etal
\cite{Su} We also show the effect of varying $\tau_s$ for the
s-wave pairing/non-magnetic structural impurity scenario, that is
that from sample to sample the minimum $\rho_0$ as a function of
cooling rate changes, but the maximum $T_c$ does not change. The
broken lines then show the expected behaviour for different
samples based on the Su \etal data assuming s-wave pairing and
non-magnetic structural impurities. Also shown is equivalent data
from experiments performed by Stalcup \etal \cite{Stalcup} It is
clear that the data from Stalcup \etal does not fit with the
expectations for s-wave pairing and non-magnetic structural
impurities. For non-s-wave pairing and/or magnetic structural
impurities both structural disorder and cooling rate induced
disorder reduce $T_c$. Thus $T_c$ is given by (\ref{eqn:AGNM})
with $\tau_N = \tau_t$. While the residual resistivity is still
determined by (\ref{eqn:resistivity_propto}).

The solid line in figure \ref{fig:d-wave_T_c_expt} represents at
fit to the data of Su \etal The fabrication of different samples
will lead to different values of $\tau_s$. This will cause a
variation from sample to sample in both the minimum value of
$\rho_0$ and the maximum value of $T_{c}$ obtainable by varying
the cooling rate. However as $T_c$ and $\rho_0$ are both functions
of only one variable ($\tau_t$) the data for all samples will lie
on a single line. Thus the broken lines in figure
\ref{fig:d-wave_T_c_expt} represent the prediction of the
behaviour of different samples based on the Su \etal data assuming
non-s-wave pairing and/or magnetic structural impurities. It is
clear that the data from Stalcup \etal is in excellent agreement
with the expectations for non-s-wave pairing.

We stress that this result is based on experiments on only two
samples. To be conclusive one would require the study of many more
samples. Further it has been argued \cite{Wosnitza_private} that
some measurements of the critical temperature and residual
resistivity in the literature \cite{Wosnitza_2003} are more
consistent with the s-wave pairing scenario (figure
\ref{fig:s-wave_T_c_expt}). Clearly, a detailed, systematic study
is required to settle this debate.

The above work is based on the (reasonable) assumption that the
structural impurities are non-magnetic. As we speculated in the
case of \bXc it may be that some novel mechanism of local moment
formation exists in the layered organic superconductors. Applying
a hydrostatic pressure or changing the anion (X) in \kX has a
dramatic effect on the ground state. For example at ambient
pressure and low temperature \kCl is a Mott--Hubbard
antiferromagnetic insulator. Applying a small pressure ($\sim
200$~bar, Ref. \onlinecite{Lefebvre}) moves \kCl into an
superconducting state with properties very similar to those of
\Brs Thus it is thought that \Br is close (in anion/pressure
space) to an antiferromagnetic phase transition.
\cite{Ross_review} A possible mechanism for the formation of local
moments in \Br is that non-magnetic impurities change the local
electronic structure by a small amount. This small local
perturbation could cause the formation of a local moment similar
to those found in the antiferromagnetic phase. A similar
suggestion was made by Kohno \etal \cite{Kohno} who considered the
competition of antiferromagnetic and superconducting ground states
in Ce$_x$Cu$_2$Si$_2$ with $x\lesssim 1$. In their scenario Ce
vacancies act as intrinsically non-magnetic impurities, but lead
to the formation of local moments. At low enough densities such
magnetic impurities will act as independent, paramagnetic spins.
As such the impurities' behaviour in a magnetic field is governed
by the Brillouin function: \cite{Ashcroft&Mermin}
\begin{widetext}
\begin{eqnarray}
M = N_Mg\mu_BJ_{i} \left[ \left(1+\frac{1}{2J_i}\right) \coth
\left(\left(1+\frac{1}{2J_i}\right) \frac{g\mu_BHJ_i}{k_BT}
\right) - \frac{1}{2J_i} \coth
\left(\frac{g\mu_BHJ_i}{2k_BT}\right) \right],
\end{eqnarray}
\end{widetext}
where $N_M$ is the total number of magnetic impurities, $J_i$ is
the total angular momentum of the impurity and $g$ is the usual
$g$-factor. For localised, non-interacting electrons it is
appropriate to take $J_i=\frac{1}{2}$ and $g\simeq2$. In which
case
\begin{eqnarray}
M = N_M\mu_B \tanh \left(\frac{\mu_BH}{k_BT}\right).
\label{eqn:mag}
\end{eqnarray}
From (\ref{eqn:mag_lifetime}) we have
\begin{eqnarray}
N_M = \frac{4}{3\pi N(0)|u_M|^2}\frac{1}{\tau_c}.
\label{eqn:No_mag_imp_tau}
\end{eqnarray}
$N(0)$ is known \cite{JM&RHM00} because for a quasi-two
dimensional metal the density of states at the Fermi level is
given by
\begin{eqnarray}
N(0) = \frac{m_c}{2\pi\hbar^2}
\end{eqnarray}
where $m_c$ is the cyclotron mass. In the presence of interactions
Luttinger's theorem \cite{Luttinger_theorem} for a Fermi liquid
ensures that \cite{JM&RHM00}
\begin{eqnarray}
N(0) = \frac{m^*}{2\pi\hbar^2}, \label{eqn:q2DDOS}
\end{eqnarray}
where $m^*$ is the effective mass, regardless of the details of
the band structure. It is known from Shubnikov--de Haas
experiments \cite{Caulfield} that, for the $\beta$ or magnetic
breakdown orbit $m^*/m_e = 6.4$ and so $N(0) = 14.9$~eV$^{-1}$unit
cell$^{-1}$spin$^{-1}$.

A more difficult problem is estimating $u_M$. We can make an
estimate because of our knowledge of the Mott--Hubbard state which
is nearby in pressure/anion space. We estimate that $u_M$ will be
of the same order as $JV$ where $J$ is the exchange coupling in
the Mott--Hubbard state and $V$ is the volume occupied by a dimer
and an anion. This is dimensionally correct and we know that in
the Mott antiferromagnetic state there is one spin per dimer. It
is estimated that $J \sim 40$~K (Ref.
\onlinecite{footnote_estimate_J})
and hence $|u_M| = 0.026$~eV\AA$^3$. A less theory-laden estimate
of $J$ can be made from the fact that the Kondo effect is not
observed in these materials. In the Kondo effect a minimum in the
resistivity occurs at the Kondo temperature, $T_K$, which is given
by \cite{Hewson}
\begin{eqnarray}
T_K=\frac{W}{k_B}\exp\left(-\frac{1}{2JN(0)}\right),
\end{eqnarray}
where $W$ is the bandwidth. For \Brc $W=2(t_1+t_2)\simeq0.23$~eV,
where $t_1$ and $t_2$ are the nearest neighbour and next nearest
neighbour hopping integrals respectively, \cite{JM&RHM00} and
$N(0)$ is given by (\ref{eqn:q2DDOS}) with $m^*/m_e=6.4$. That the
Kondo effect is not observed implies that $T_K<T_c<T_{c0}<12$~K
from the fit in figures \ref{fig:s-wave_T_c_expt} and
\ref{fig:d-wave_T_c_expt}. This implies that $J<155$~K and thus
that $|u_m|^2<0.4$~eV\AA$^3$. However, while the Kondo temperature
is defined for a single impurity, the Kondo minimum will not be
observable unless there is a sufficiently large number of
impurities (typically a few percent \cite{Hewson}).

Substituting (\ref{eqn:linearisedAG}) into
(\ref{eqn:No_mag_imp_tau}) we find that
\begin{eqnarray}
N_M = \frac{32k_B}{3\pi^2\hbar N(0)|u_M|^2} (T_{c0} - T_c).
\label{eqn:No_mag_imp_Tc}
\end{eqnarray}
For example, Su \etal \cite{Su} report a maximum variation in the
critical temperature of $T_{c0} - T_c = 0.58$~K, which leads to,
as a lower bound (based on $J \sim 155$~K), $N_M \gtrsim 0.03$
impurities per unit cell. For our best guess ($J \sim 40$~K) we
find $N_M \gtrsim 0.50$ impurities per unit cell. This should be
sufficient to observe a Kondo minimum and thus the Kondo effect
places a limit on the number of impurities.

Substituting (\ref{eqn:No_mag_imp_Tc}) into (\ref{eqn:mag}) we
find that,
\begin{eqnarray}
\frac{M}{\mu_B} = \frac{32k_B}{3\pi^2\hbar N(0)|u_M|^2}
\left(T_{c0} - T_c\right) \tanh \left(\frac{\mu_BH}{k_BT}\right).
\end{eqnarray}

\begin{figure}
    \centering
    \epsfig{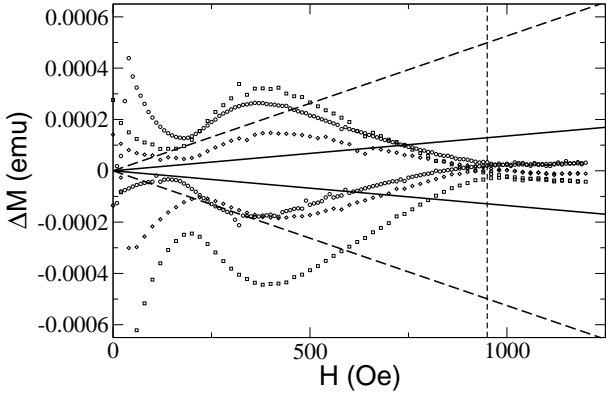}
    \caption{The cooling rate dependence of the magnetisation of
    \Brs We plot the difference in the magnetisation of the same sample
    when it is cooled at
    10 K/min and when it is cooled at 0.5 K/min (circles) measured
    by Taniguchi and Kanoda. \cite{Taniguchi&Kanoda} Also shown is
    the difference in the magnetisation for the same sample when
    it had been annealed at 70~K for 12 hours and when it was
    cooled at 0.5 K/min (diamonds) and the difference in
    magnetisation between when the sample was annealed and when it was
    cooled at 10 K/min (squares). All sets of data were
    taken at $T=7\textrm{~K}<T_c$.
    The solid lines are the
    calculated lower bound on the
    change in the magnetisation at $T=7$~K due to paramagnetic impurities which
    produce a 0.25~K change in $T_c$ which is the estimated change
    in $T_c$ between the sample cooled at 10~K/min and the sample
    cooled at 0.5~K/min based on the observed cooling
    rate dependence of this sample. \cite{Taniguchi_private} This
    lower bound is required to ensure the Kondo temperature,
    $T_K<T_c$ and thus to be consistent with the fact that the
    Kondo effect is not observed in \Brs The long dashed lines
    represent the predicted magnetisation assuming that the
    interaction energy
    of the magnetic impurities is the same as the observed
    antiferromagnetic exchange interaction in the insulating phase of \kCl (i.e., $J\sim40$~K).
    The vertical dashed line indicates the
    irreversibility line at 7~K, $H_{\textrm{ir}}(T=7\textrm{ K})$, as measured by Taniguchi and Kanoda
    \cite{Taniguchi&Kanoda} in the same experiment. Thus we see that
    for $H<H_{\textrm{ir}}(T=7\textrm{ K})$ (left of the dashed line)
    the non-trivial vortex dynamics of the system cause a complicated variation in the
    magnetisation, which we do not discuss here. However, for
    $H>H_{\textrm{ir}}(T=7\textrm{ K})$ (right of the dashed line) the
    measured difference in the magnetisation is less than that required
    by the Brillouin function. Therefore these measurements suggest that
    no paramagnetic impurities are induced by varying the cooling rate of this sample.
    But this conclusion requires that the moments are not screened by supercurrents.} \label{fig:Taniguchi}
\end{figure}

Two studies of the variation in magnetisation with cooling rate in
\Br have been conducted. \cite{Aburto,Taniguchi&Kanoda} Both
studies were primarily concerned with the weak field limit, but
surprisingly even these results may tell us something about the
presence of magnetic impurities. Taniguchi and Kanoda
\cite{Taniguchi&Kanoda} measured $M(H)$ at $T=7$~K. They found an
interesting weak field dependence (presumably this is due to
vortex dynamics as it disappears when the irreversibility line is
reached, but we will not discuss this here). Above the
irreversibility line they found that the change in $M$ with
cooling rate is only weakly dependant on $H$. (Results were
reported up to $H=1200$~Oe.) Based on the observed cooling rate
dependence of $T_c$ in this sample \cite{Taniguchi_private} we
estimate that the variation in $T_c$ between when the sample is
cooled at 10~K/min and when the sample is cooled at 0.5~K/min is
0.25~K. This leads to the conclusion that the difference in the
magnetisation of the two samples due to the magnetic impurities
(required in the s-wave scenario) would be $1.3\times10^{-4}$~emu
at H=1200~Oe and T=7~K (based on our lower bound from the Kondo
effect, $J=155$~K). This is well within the resolution of the
experiment (in fact this contribution would dominate the observed
magnetisation) and \emph{is not observed} (see figure
\ref{fig:Taniguchi}). Thus the experiments of Taniguchi and Kanoda
are inconsistent with the hypothesis that cooling rate induced
disorder creates paramagnetic impurities. (However, it is possible
that paramagnetic impurities are present in the sample and that
there presence is screened by the superconducting state.) We
therefore suggest that there is non-s-wave pairing in \Br and that
varying the cooling rate induces non-magnetic disorder which
causes the variation in both $T_c$ and $\rho_0$. Again we stress
that because the is little data above the irreversibility line,
$H_{\textrm{ir}}$, and no data outside the superconducting state,
further careful systematic experiments are required preferably in
the normal state.

Two groups have investigated anomalies in heat capacity
\cite{Saito,Sato} and thermal expansion \cite{MullerGlass} at
$T\sim 80$~K in $\kappa$-(BEDT-TTF)$_2$X for X = Cu[N(CN)$_2$]Cl,
(Refs. \onlinecite{Sato} and \onlinecite{MullerGlass})
Cu[N(CN)$_2$]Br (Refs. \onlinecite{Saito}, \onlinecite{Sato} and
\onlinecite{MullerGlass}) and Cu(NCS)$_2$ (Ref.
\onlinecite{MullerGlass}). Both groups concluded that the
anomalies are due to a transition in which disorder becomes frozen
into the orientational degrees of freedom in the terminal ethylene
groups of the BEDT-TTF molecules. This ethylene ordering
transition provides a natural explanation for the observed cooling
rate dependence of the residual resistivity of
$\kappa$-(BEDT-TTF)$_2$X. However one should note that such an
ethylene ordering transition would result in intrinsically
non-magnetic impurities and is therefore strong evidence in
support of our suggestion that the cooling rate induced disorder
is non-magnetic in nature.

Terminal ethylene group disorder in \kX is rather similar to the
anion disorder observed in the Bechgaard salts. In both
(TMTSF)$_2$ClO$_4$ and (TMTSF)$_2$RuO$_4$ the anions can occupy
two inequivalent orientations. Fast cooling leads to partially
disordered domains, the size of the domains has been shown to be
proportional to the cooling rate. \cite{Pouget} As mentioned in
section \ref{sect:other_SCs} varying the cooling rate can lead to
a reduction in $T_c$ and even the complete suppression of
superconductivity in favour of a spin density wave. Also note that
the anion ordering temperature, $T_{AO}$, is highly dependent on
which anion is considered. For X = ClO$_4$, $T_{AO} \sim 24$~K,
for X = ReO$_4$, $T_{AO} \sim 170$~K and for X = PF$_6$ no anion
ordering transition is observed. \cite{Tsobnang} The nature of the
anion order also differs for X = ClO$_4$ and X = ReO$_4$ (Ref.
\onlinecite{Tsobnang}). A similar disordering transition is
observed \cite{Akutsu} in the organic conductors (DMET)$_2$BF$_4$
and (DMET)$_2$ClO$_4$.

Of the salts considered here, a variation in $T_c$ with cooling
rate had only been observed in \Br to date. If our hypothesis that
the variation in $T_c$ with cooling rate is due to cooling rate
induced disorder which in turn is due to the ethylene ordering
transition in the terminal ethylene groups is correct then one
would also expect a variation in $T_c$ with cooling rate in \NCS
as the ethylene ordering transition has been observed in this
compound. \cite{MullerGlass,footnote_coolingrate_NCS} An ethylene
ordering transition has also been observed in
$\kappa$-(BEDT-TTF)$_2$Cu[N(CN)$_2$]Cl (Ref.
\onlinecite{MullerGlass}). However, this compound only becomes
superconducting under pressure and it is not known what effect
pressure has on the disordered ethylene state. Clearly the
dependence of $T_c$ on cooling rate is in need of further
investigation. It may be of interest to investigate the effect of
pressure on the ethylene ordering transition, particularly with
reference to $\kappa$-(BEDT-TTF)$_2$Cu[N(CN)$_2$]Cl and cooling
rate dependence of the N\'{e}el temperature.

In light of the variation of $T_c$ with cooling rate it is
important that in experiments on the $\kappa$-(BEDT-TTF)$_2$X
salts the cooling rate is reported regardless of whether or not it
is varied. Results for $T\lesssim80$~K lose much of their
significance if the cooling rate is not known.

Work by Taniguchi \etal \cite{Taniguchi,Taniguchi03} has raised
the possibility of inhomogeneous phase coexistence between
antiferromagnetism and superconductivity in deuterated \Brs There
is no evidence of phase coexistence in fully hydrogenated \Br so
phase coexistence can be ruled out as the cause of the suppression
of $T_c$ in the hydrogenated compound, which we consider here.
Further varying the cooling rate of deuterated \Br and \kCl offers
the possibility of studying the Mott transition in the presence of
disorder with fine experimental control over the level of disorder
in the sample and of varying the level of disorder within a single
sample.

\section{Interlayer transport theory}\label{sect:trans_theory}

The residual resistivity for interlayer transport in a layered
Fermi liquid is given by (see, for example, Ref.
\onlinecite{Ross&Moses})
\begin{eqnarray}
\rho_0 = \frac{\pi\hbar^4}{2e^2m^*ct_\perp^2}\frac{1}{\tau_t}
\label{eqn:interlayer_resistivity}
\end{eqnarray}
where $c$ is the interlayer spacing, $m^*$ is the effective
quasiparticle mass and $t_\perp$ is the interlayer hopping
integral. Thus the assumption that $\rho_0\propto 1/\tau_t$
(\ref{eqn:resistivity_propto}) is justified.

Substituting (\ref{eqn:interlayer_resistivity}) into
(\ref{eqn:linearisedAG}) we find that
\begin{eqnarray}
T_c = T_{c0} - \frac{e^2m^*ct_\perp^2}{4k_B\hbar^3} \rho_0.
\label{eqn:prediction}
\end{eqnarray}
Thus from our fit to the data of Su \etal \cite{Su} (shown in
figure \ref{fig:d-wave_T_c_expt}) we have, for \Brc
\begin{eqnarray}
T_{c0} = 11.7~ \textrm{K}
\end{eqnarray}
and
\begin{eqnarray}
\frac{e^2m^*ct_\perp^2}{4k_B\hbar^3} = 0.9~\Omega\textrm{cm}.
\end{eqnarray}
Taking $m^* = 6.4m_e$ (Refs. \onlinecite{Weiss} and
\onlinecite{JM&RHM00}) and $c = 30.016$~\AA ~(Ref.
\onlinecite{Ishiguro}) we have $t_\perp = 0.022$~meV. However, we
note that $m^*$ was determined for the $\beta$-sheet (that is the
magnetic breakdown orbit) only whereas here we are considering an
effective one band model. Nevertheless this value is in excellent
agreement with an independent determination of $t_\perp$ from
angular-dependent magnetoresistance (AMRO) experiments. Although
$t_\perp$ has not been measured experimentally in \Brc for \kI
$t_\perp \approx 0.016$~meV (Ref. \onlinecite{Wosnitza_t_c_kI})
and for \NCS $t_\perp \approx 0.04$~meV (Ref.
\onlinecite{Singleton_t_c_NCS}).

For \bIBr (see figure \ref{figure:IBr2}) we find that $T_{c0} =
3.0$~K. Tokumoto \etal \cite{Tokumoto} reported that the room
temperature resistivity of their samples was $\rho(295) = (5.0 \pm
2.5) \times 10^{-2}~\Omega$cm. Therefore
\begin{eqnarray}
\frac{e^2m^*ct_\perp^2}{4k_B\hbar^3} = 40\pm20~\Omega\textrm{cm}.
\end{eqnarray}
Taking $m^* = 4.2 m_e$ (Ref. \onlinecite{Ishiguro}) and $c =
15.291$~\AA ~(Ref. \onlinecite{Ishiguro}) we have $t_\perp =
0.26\pm0.07$~meV. Note that this is an order of magnitude larger
than for \Brs However, this value is also in agreement with
previous estimates from de Haas--van Alphen experiments. Wosnitza
\etal \cite{Wosnitza_ratio_kIBr} showed that for \bIBrc
$t_\perp/E_F \approx 1/280$ they also found that $k_F \sim
3.46\times10^9$~m$^{-1}$. Therefore taking $m^* = 4.2 m_e$ (Ref.
\onlinecite{Ishiguro}) again and assuming a cylindrical Fermi
surface
\begin{eqnarray}
E_F \cong \frac{\hbar^2k_F^2}{2m^*}
\end{eqnarray}
one finds that $t_\perp \approx 0.35$~meV in excellent agreement
with our result.

The agreement between $t_\perp$ calculated from our fits via
equation (\ref{eqn:interlayer_resistivity}) and the values found
from AMRO experiments for both \bIBr and \Br is further evidence
that in these compounds $T_c$ is suppressed by the AG mechanism
and not by weak localisation.

It has recently been shown \cite{betaHbetaL} that the observed
variation of $T_c$ and $\rho_0$ for alloy
$\beta$-(BEDT-TTF)$_2$(I$_3$)$_{1-x}$(IBr$_2$)$_x$ for small $x$
predicted by (\ref{eqn:prediction}) is consistent with the
observations of Tokumoto \etal \cite{Tokumoto} Note that this
theory has no free parameters once the $T_{c0}$ (this work and
Forro \etal \cite{Forro}) and $t_\perp$ (AMRO experiments
\cite{Wosnitza_t_c_kI}) have been determined.

The agreement between our calculated values of $t_\perp$ and those
measured in AMRO experiments indicates that if there is an
$\textrm{s}+\textrm{n}$ state then the s-wave component
($\cos(\varphi))\Delta_s$) is small (see section \ref{sect:AG}).
(Or more strictly that $\alpha$ is small, c.f. equation
(\ref{eqn:AGsin}).) It therefore appears unlikely that the layered
organics are $\textrm{s}+\textrm{n}$ superconductors.

\section{Discussion}

This study of the effects of disorder on the layered organic
crystals \bX and \kX has shown that disorder has the potential to
differentiate between s-wave and non-s-wave pairing states. But,
more experiments are needed. This is largely because none of the
experiments that we have discussed in this paper were designed to
study the pairing symmetry. In this section we will explore what
the unresolved issues are and how they could be resolved.

\subsection{Sample variation}

Perhaps the simplest test for unconventional superconductivity is
to study the variations in the superconducting critical
temperature reported in the literature. Crystal growers go to
great lengths to avoid the inclusion of magnetic impurities, but
the inclusion of non-magnetic impurities\footnote{Vacancies act in
exactly the same way as non-magnetic impurities. \cite{Shoenberg}}
is harder to avoid. For example the first reports of
superconductivity in Sr$_2$RuO$_4$, which is widely considered to
have an unconventional (triplet) pairing symmetry, indicated that
$T_c = 0.93$~K (Ref. \onlinecite{SCinSRO}). However, sample
quality was rapidly improved and it is now believed that the
maximum critical temperature, $T_{c0}=1.5$~K (Ref.
\onlinecite{Mackenzie_Tc}) has been achieved. Thus, for
Sr$_2$RuO$_4$, $T_c$ has increased by over $50\%$ since the first
report of superconductivity. In contrast, consider MgB$_2$. The
first report \cite{Nagamatsu} of superconductivity quoted
$T_c=39$~K. No significant increase in $T_c$ has been reported
thus far. This is evidence for s-wave pairing in MgB$_2$. Further,
doping MgB$_2$ with U does not significantly alter $T_c$ (doping
with 1 wt\% U reduces $T_c$ by $< 0.5\%$, Ref.
\onlinecite{Silver}). This is in agreement with the emerging
consensus that MgB$_2$ is a strong coupling s-wave superconductor.
\cite{Choi} (For a fuller discussion of the effects of disorder in
MgB$_2$ see Ref. \onlinecite{Erwin}.)

The initial reports of superconductivity in \Br quote $T_c =
10.8$~K (Ref. \onlinecite{Kwok}). While we have shown that $T_{c0}
= 11.7$~K. \NCS also shows wide variation in $T_c$ from sample to
sample some authors have reported $T_c$ as low as 8.7~K (Ref.
\onlinecite{Ito}), while other studies have found that $T_c =
9.3$~K (Ref. \onlinecite{Wosnitza_2003}). One complication arises
from the variety of definitions used to determine $T_c$. Taking a
resistivity measurement as an example, the $T_c$ can be defined in
a variety of ways: (i) the temperature at which $\rho$ first
begins to deviate from the Fermi liquid form ($\rho(T) = \rho_0
+AT^2$), (ii) the highest temperature at which $\rho(T) =0$, or
(iii) the midpoint of the transition, i.e., the temperature at
which the $\rho(T)$ is $50\%$ of the Fermi liquid value. For
example, definitions (i) and (ii) give a difference of $\sim 1$~K
for the data reported by Stalcup \etal \cite{Stalcup} about the
value $T_c = 11.6$~K (defined by method (iii), which we use
throughout this paper). The large variations in $T_c$ noted above
($8\%$ for \Br and 7$\%$ for $\kappa$-(BEDT-TTF)$_2$Cu(NCS)$_2$)
are probably too large to be explained by subtle variations in the
definition of $T_c$ and are therefore unlikely to occur for s-wave
pairing although this is far from conclusive.

\bI shows a strong variation in $T_c$. In the $\beta_H$ phase
\cite{Ishiguro} Kahlich \etal \cite{Kahlich} reported that $T_c$
varied between 4.5~K and 7~K depending on which sample they
measured. This represents a $36\%$ variation in $T_c$. This is
also suggestive of non-s-wave pairing.

The wide variation in $T_c$ from sample to sample is something
that great care should be taken over in experiments designed to
study the isotope effect. In particular any such experiments need
to demonstrate that crystals that are nominally identical do
indeed have a highly reproducible $T_c$. If this is not possible
then the $T_c$ variation within nominally identical samples needs
to be carefully accounted for. For example, by studying the sample
dependence of the residual resistivity across a range of nominally
identical samples and using this to calibrate the impurity
dependence of the various isotopes.

\subsection{Measurement of the scattering time}

Disorder would be a much more powerful probe if there existed a
method by which the scattering time could be measured directly.
The most obvious techniques for this are Shubnikov--de Haas and de
Haas--van Alphen experiments. These quantum oscillation
experiments measure the quasiparticle lifetime via the Dingle
temperature, $T_D$. However, the lifetime determined by quantum
oscillation experiments, $\tau_q$, is not the same as the
transport lifetime, $\tau_t$ (Refs. \onlinecite{Shoenberg} and
\onlinecite{BenThesis}). Even in the best experiments on elemental
metals, it is not at all unusual for $\tau_t$ to be 10 or even 100
times larger than $\tau_q$ (Ref. \onlinecite{Shoenberg52}). In
particular $T_D$ and hence $\tau_q$ are known to be very sensitive
to the mechanical state of the sample. A slight deformation caused
by, for example, handling the sample can lead to dramatic increase
in $T_D$ (decrease in $\tau_q$), whilst hardly affecting the
electrical resistivity ($\rho_0\propto1/\tau_t$). Given the large
compressibility of the layered organic superconductors $\tau_q$ is
unlikely to be the same as $\tau_t$.

In its immediate location a dislocation acts just like a line of
point defects and thus contributes equally to both transport and
quantum oscillation experiments. However, the long-range strain
field produced by a dislocation only produces very small angle
scattering (as the electron wavelength is smaller than the
characteristic length scale of the dislocations). Therefore the
long-range strain field contributes negligibly to the transport
lifetime but can strongly suppress $\tau_q$ even at relatively low
dislocation densities.

A sample which is nominally a single crystal is in fact made up of
a large number of grains. One can think of this mosaic structure
of grains as a certain pattern of dislocations. In this way it is
clear that mosaic structure causes highly anisotropic scattering
and thus leads to the suppression of $\tau_q$.

Many previous authors have pointed out the difference in the
transport and quantum lifetimes. However, Hill \cite{Hill} noted a
similar difference between the lifetime observed in cyclotron
resonance experiments, $\tau_\textrm{cr}$, and the quantum
lifetime. It is therefore interesting to compare the lifetimes
from cyclotron resonance and quantum oscillation experiments with
the transport lifetime determined from the linearised AG equation
(\ref{eqn:linearisedAG}) and the value of $T_{c0}$ found from the
fit to experiment, $\tau_t^\textrm{AG}$, (see Table
\ref{tab:lifetimes}).

\begin{table}
\caption{\label{tab:lifetimes}Comparison of the
transport/Abrikosov--Gorkov, cyclotron resonance and quantum
oscillation quasiparticle lifetimes ($\tau_t^\textrm{AG}$,
$\tau_\textrm{cr}$ and $\tau_q$ respectively).  As $\tau$ is
clearly a highly sample dependent property this table is not
intended to report universal results but is indicative of general
trends. NS indicates a non-superconducting compound for which
$\tau_t^\textrm{AG}$ cannot be determined. The reported
$\tau_t^\textrm{AG}$ is based on the samples used for the
experiments discussed in this paper (or in Ref.
\onlinecite{BenThesis} in the case of ZrZn$_2$). We have
abbreviated BEDT-TTF to ET in this table.}
\begin{ruledtabular}
\begin{tabular}{lccc}
Material&$\tau_t^\textrm{AG}$ (ps)&$\tau_\textrm{cr}$ (ps)&$\tau_q$ (ps)\\
\hline \vspace*{-9pt} \\
$\kappa$-(ET)$_2$Cu[N(CN)$_2$]Br & 2.5-20 & ? & 0.5-0.6 \cite{Stalcup}\\
$\beta$-(ET)$_2$IBr$_2$ & 0.6-1.8 & ? & 1.5 \cite{Kartsovnik_2_Dingle_temps} \\
$\beta$-(ET)$_2$I$_3$ & ? & ? & 2.4 \cite{Kartsovnik} \\
$\theta$-(ET)$_2$I$_3$ & ? & 15-36 \cite{Oshima} & 0.6-1.5 \cite{Tamura} \\
$\alpha$-(ET)$_2$KHg(NCS)$_4$ & ?\footnote{$\alpha$-(ET)$_2$KHg(NCS)$_4$ is only superconducting under
pressure\cite{Andres}} & 15 \cite{Kovalev} & 0.3-0.5 \cite{Sasaki} \\
$\alpha$-(ET)$_2$NH$_4$Hg(NCS)$_4$ & ? & 40 \cite{Hill} & 2 \cite{Wosnitza} \\
(ET)$_2$Br(DIA) & NS & 4.6-5.5 \cite{Oshima} & 1.7 \cite{Oshima} \\
(ET)$_3$Cl(DFBIB) & NS & 5.6 \cite{Oshima} & 1.7 \cite{Oshima} \\
Sr$_2$RuO$_4$ & 6-38\footnote{The sample measured by Hill
\etal\cite{HillSRO} had $T_c = 1.44$~K (for which the AG formula
gives $\tau_t^\textrm{AG} = 37.9$~ps based on $T_{c0}=1.52$~K, the
value found from fitting the AG formula to the data of Mackenzie
\etal \cite{Mackenzie_Tc}). $\tau_t^\textrm{AG}=6.25$~ps based on
$T_c = 1$~K, the value reported in the de Haas-van Alphen
experiments. \cite{Mackenzie}}
& 10-40 \cite{Hill_private} & 1.8 \cite{Mackenzie} \\
ZrZn$_2$ & $\sim6$ \cite{BenThesis} & ? & 0.3 \cite{Yates} \\
\end{tabular}
\end{ruledtabular}
\end{table}

We see that $\tau_t^\textrm{AG} \sim \tau_{cr}$ across a broad
range of (BEDT-TTF)$_2$X salts, while $\tau_q$ is consistently an
order of magnitude smaller. This suggests that scattering events
are not the dominant contribution to $T_D$ (c.f., Singleton \etal
\cite{Singleton}). It presents the intriguing possibility that
cyclotron resonance experiments could be used to probe the
quasiparticle lifetime and thus directly compare the experimental
$T_c$ with the predictions of the AG equation. Indeed cyclotron
resonances have already been observed \cite{HillSRO,Hill_private}
in Sr$_2$RuO$_4$. The observed cyclotron resonance lifetime is
larger than the observed lifetime in de Haas-van Alphen
experiments, but this may be partly explained by the much higher
$T_c$ of the sample used for the cyclotron resonance experiments.
Excellent agreement is found between the measured cyclotron
lifetime and the lifetime calculated from the AG formula. Clearly,
a systematic study of how the cyclotron resonance lifetime (and
indeed the quantum oscillation lifetime) varies with $T_c$ is
needed. \Sr would be an ideal material for such experiments as the
AG formula is seen to be obeyed, \cite{Mackenzie_Tc} and good
quality quantum oscillation \cite{Mackenzie} and cyclotron
resonance experiments \cite{HillSRO} can be performed.
Alternatively the AG behaviour of \Br would make it an excellent
material for such a experiment. This is particularly elegant as
the cooling rate can be used to vary the disorder and hence the
scattering lifetime, so the experiment could be performed on a
single sample. Measurements of the variation of the Dingle
temperature with cooling rate have already been made.
\cite{Stalcup}

Kartsovnik, Grigoriev and coworkers
\cite{Kartsovnik_2_Dingle_temps,Grigoriev} have also investigated
the relationship between the quasiparticle lifetimes caused by
solely microscopic scattering events, and the lifetime extracted
from the Dingle temperature which also contains the effects of
macroscopic inhomogeneities. They have shown that the slow
oscillations observed in quantum oscillation experiments on
quasi-two dimensional metals are damped by a modified Dingle
temperature, $T_D^*$, which is not effected by macroscopic
inhomogeneities. For experiments performed on \bIBr they found an
order of magnitude difference between $\tau_q$ (1.5~ps) and the
lifetime derived from $T_D^*$, $\tau_q^*$ (8.1~ps).

The Fermi velocity, $v_F$, for both the $\beta$ and $\kappa$
polymorphs is typically $v_F\sim10^5$~ms$^{-1}$ (see section
\ref{sect:trans_theory} and Ref. \onlinecite{Kovalev_vF}). And we
have shown here that a quasiparticle lifetime of the order
$\tau_t^{AG}\sim0.1$~ps is required to completely suppress
superconductivity. Thus the mean free path, $l=v_F\tau_t$, is
typically $l\gtrsim10\textrm{ nm}$ (c.f. Ref.
\onlinecite{Dressel}). The interlayer coherence length,
$\xi_\parallel$, is typically a few nm (c.f. Ref.
\onlinecite{Ishiguro}). Thus these materials are in the clean
limit even when superconductivity is completely suppressed by
disorder. This is further confirmation that the AG mechanism is
responsible for the suppression of superconductivity in these
materials.

We will conclude this section by outlining a series of experiments
that could determine if the disorder in the layered organic
superconductors is due to scattering from localised moments or
potential scattering. These experiments therefore have the
potential to rule out s-wave pairing.

\subsection{Identification of the pairing symmetry}

Comparatively little attention has been focused on the pairing
symmetry of \bX so we will start by considering this crystal
structure. All of the methods of creating disorder considered in
this paper (namely fast electron irradiation, alloying anions and
accidental disorder from the fabrication process) should be
revisited and studied in more depth. Both figures
\ref{figure:IBr2} and \ref{figure:I3} need more data points.
Therefore it is most important that the entire AG is mapped out.
In particular it is important to observe the complete suppression
of superconductivity by very small amounts of disorder that is a
unique feature of the AG formalism. Careful observation of the
entire AG curve is required to rule out other mechanisms for the
suppression of $T_c$ such as weak localisation, interband
scattering, changes in the pairing interaction or the macroscopic
coexistence of superconducting and non-superconducting phases. All
of these mechanism for the suppression of $T_c$ produce markedly
different relationships between $T_c$ and $\rho_0$ and thus would
be ruled out by the observation of the entire AG curve and in
particular the complete suppression of $T_c$ by moderate amounts
of disorder which is not caused by any of the other mechanisms for
$T_c$ suppression. Forro \etal \cite{Forro} did not measure the
resistivity of their irradiated samples. It is important to know
the resistivity for several reasons: (i) it allows for easy
comparison with other techniques, in particular it allows a
consistent definition of $T_c$ to be used, (ii) it provides a
check on the estimation of the number of defects produced and
(iii) it allows for the calculation of $t_\perp$ and thus for a
further check that AG theory is indeed relevant. All of these
methods should also be applied to \kXs

The next step is to discover whether any of the methods for
producing impurities create magnetic scatterers. One way to do
this is to measure the magnetisation as Taniguchi and Kanoda
\cite{Taniguchi&Kanoda} have for cooling rate induced disorder in
\Brs However, this experiment should be repeated in the metallic
state. This suggests that paramagnetic impurities are not induced
by varying the cooling rate of \Brs

Here we will consider alternative experiments which could be used
to search for magnetic impurities. We will describe these
experiments in the context of cooling rate induced disorder in
\Brs However, the generalisation of these experiments to the other
methods of producing disorder is straightforward. Cooling rate
induced disorder experiments are particularly elegant as the level
of disorder can be controlled within a single sample. This reduces
systematic errors, for example, by far the largest source of error
in measuring $\rho_0$ comes from measuring the samples dimensions,
such errors cancel in cooling rate induced disorder experiments.

Muon spin relaxation ($\mu$SR) experiments are capable of
detecting localised magnetic moments. \cite{Dalmas} If local
moments are produced, then the muon spin relaxation rate would
vary as a function of cooling rate. Clearly the muon spin
relaxation rate is changed by the superconducting state. As $T_c$
and presumably $H_{c2}$ are changed by the cooling rate it is
important that these experiments be done in the
non-superconducting state, either above $T_c$ or above $H_{c2}$.
As the ethylene ordering transition occurs at $T \sim 80$~K and
$T_c \sim 10$~K any local moments should be well formed several
Kelvin above $T_c$.

Nuclear quadrupole resonance (NQR) experiments have been used to
observe the formation of local moments in La$_{2-x}$Sr$_x$CuO$_4$
(LSCO) for $x=0.06$ (Ref. \onlinecite{Julien}). As perviously
discussed NMR measurements have observed localised moments induced
by Zn impurities in YBCO. \cite{Mahajan} Therefore studying the
change in $1/T_1$ with cooling rate in \Br could determine whether
or not local moments are formed. The change in $1/T_1$ as a
function of cooling rate has been measured in 98\% deuterated \Brs
No change in $1/T_1$ was observed until below 30~K, in particular
$1/T_1$ is independent of cooling rate near 80~K where the
ethylene ordering transition occurs. \cite{Kawamoto} However, fast
cooling of deuterated \Br drives the ground state from
superconductivity to an antiferromagnetic Mott insulator
\cite{Taniguchi,Taniguchi03} (which causes the observed difference
in $1/T_1$ below 30~K). Therefore this observation does not rule
out the possibility of local moments in hydrogenated \Brs Wang
\etal \cite{Wang} carried out an electron spin resonance (ESR)
experiment on \NCSs 
Wang \etal saw no signal attributable to Cu(II) species at any
temperature although they do not comment on other sources of
magnetic impurities. Therefore it is reasonable to hope that
further ESR studies may shed some light on the issue of magnetic
impurities.

The techniques, outlined here, for using intrinsically
non-magnetic disorder to probe the superconducting state are
clearly more general than the context of \bX and \kX that we have
examined here. Disorder has already been used to study
Sr$_2$RuO$_4$ (Ref. \onlinecite{Mackenzie_Tc}) (although we should
note that no experiments have been performed to rule out magnetic
impurity formation in this material). Similar results for UPt$_3$
(Ref. \onlinecite{Duijn}) appear to have gone largely unnoticed.
Clearly more careful analysis of this work is required. These
methods could also be extended to other heavy fermion
superconductors. There are several other quasi-two-dimensional
organic superconductors (such as $\lambda$-(BETS)$_2$X,
$\theta$-(BEDT-TTF)$_2$X and $\beta''$-(BEDT-TTF)$_2$X) which may
be unconventional superconductors. Disorder would appear to be a
powerful tool for the investigation of the superconducting state
in these materials.

But, the study of disorder, perhaps, is most powerful when used to
identify s-wave pairing. An excellent example from the recent past
is the high temperature superconductor, MgB$_2$, which appears to
be a phonon mediated s-wave superconductor. \cite{Choi} This could
be confirmed by careful study of the effects of disorder and
showing that disorder can be introduced with only a small change
in $T_c$ (c.f. Ref. \onlinecite{Erwin}). This could also be
applied to other superconductors which are suspected of being
s-wave, in particular superconductors suspected of having
anisotropic s-wave order parameters, such as the borocarbides.
\cite{Martinez-Samper}

\section{Conclusions}

We have considered the effect of impurities and disorder on the
superconducting critical temperature in \bX and \kXs We have shown
that various sources of disorder (alloying anions, \cite{Tokumoto}
fast electron irradiation, \cite{Forro} disorder accidentally
produced during fabrication, \cite{Shegolev} and cooling rate
induced disorder \cite{Su,Stalcup}) lead to a suppression of $T_c$
that is well described by the Abrikosov--Gorkov formula. This is
confirmed not only by the excellent fit to a theory with only two
free parameters, but also by the excellent agreement between the
value of the interlayer hopping integral, $t_\perp$, calculated
from this fit and the value of $t_\perp$ found from AMRO
experiments. This makes a pairing state with a superposition of
s-wave and non-s-wave components extremely unlikely. Although such
and $\textrm{s}+\textrm{n}$ state cannot be strictly ruled out,
the s-wave part of the wavefunction must be very small and the
coherence between the s-wave and non-s-wave parts of the
wavefunction must be completely rigid
($\alpha(\varphi(\tau))=\alpha \ll 1$). The agreement between the
measured and calculated values of $t_\perp$ effectively leaves
$T_{c0}$ as the only free parameter in the theory. In practice one
has very little choice over the value of $T_{c0}$, so the
agreement with experiment is found from an essentially parameter
free theory. The AG formula describes the suppression of $T_c$ by
magnetic impurities in singlet superconductors, including s-wave
superconductors. However $T_c$ is suppressed in exactly the same
way by non-magnetic impurities in a non-s-wave superconductor. We
therefore have shown that there are only two scenarios consistent
with the current state of experimental knowledge. We summarise
these scenarios below. The task is now to discover whether the
impurities are magnetic or non-magnetic.

\emph{Scenario 1: d-wave pairing.} If the disorder induced by all
of the four methods considered in this paper is, as seems most
likely, non-magnetic then the pairing state cannot be s-wave.
Triplet pairing is ruled out by the combination of the three
experiments discussed in section \ref{sect:intro}.
\cite{de_Soto,Zuo,Murata} Therefore we know that the angular
momentum, $l$, of the Cooper pairs is even. If the disorder does
turn out to be non-magnetic then this implies that $l\geq2$. In
which case Occam's razor suggests that d-wave pairing is realised
in both \bX and \kXs

\emph{Scenario 2: a novel mechanism for the formation of local
magnetic moments.} 
Given the proximity of \bX and \kX it the Mott-Hubbard
antiferromagnetic state in anion/pressure space, it is possible
that disorder induces local magnetic moments. Further Taniguchi
\etal \cite{Taniguchi03,Taniguchi} have suggested that varying the
cooling rate can to the macroscopic coexistence of
superconductivity in deuterated \Brs Although there is no evidence
for anything but a spatially uniform superconducting state in the
hydrogenated compound, \cite{Taniguchi03,Taniguchi} which we have
considered here, these experiments would not detect isolated
magnetic impurities. On the other hand we have shown here that the
work of Taniguchi and Kanoda \cite{Taniguchi&Kanoda} is
inconsistent with the theory that disorder modulates the local
electronic structure and thus moves single sites or small clusters
of sites into a state, analogous to the Mott--Hubbard insulating
state with localised electrons, which can act as magnetic point
scatterers. However, only a little data was reported in the
relevant magnetic field range so further work is needed to
rigourously test this scenario.

We have suggested experiments to differentiate between these
scenarios. Such experiments would either discover a novel
mechanism for the production of localised magnetic moments or
determine that the superconducting order parameter is d-wave in
\bX and \kXs

\section*{Acknowledgements}

This work was stimulated by discussions with Russ Giannetta and
Eugene Demler. We would like to thank Jim Brooks, David Graf,
Stephen Hill, Kazushi Kanoda, Hiromi Taniguchi and Fulin Zuo for
providing us with their data and for helpful discussions. We would
like to thank Werner Biberacher, Mark Kartsovnik, Igor Mazin,
Jeremy O'Brien, Claude Pasquier, David Singh, John Singleton and
Jochen Wosnitza for drawing our attention to several important
results and for their comments on an earlier version of this
manuscript. This work was supported by the Australian Research
Council.

\bibliographystyle{apsrev}
\bibliography{C:/central_latex_files/brisbane}

\begin{thebibliography}{172}
\bibitem[{foo({\natexlab{a}})}]{footnote_estimate_J}
\bibinfo{note}{The estimate $J \sim 40$~K (Ref. \onlinecite{Ross_review}) is
  based on the the antiferromagnetic state of \kClc which is the material
  closest to \Br in the temperature-pressure/anion phase diagram of these
  materials. \cite{Kanoda,Ross_review} It should be noted however that in the
  spin liquid compound $\kappa$-(BEDT-TTF)$_2$Cu$_2$(CN)$_3$ estimates
  \cite{Shimizu} based on the Heisenberg model on an isotropic triangular
  lattice give $J\sim250$~K.}
\end{thebibliography}
\end{document}